\documentclass[twocolumn,showpacs,preprintnumbers,amsmath,amssymb]{revtex4}
\usepackage{epsfig}
\begin{document}

\title{Observation of a Distribution of Internal Transverse Magnetic Fields in a Mn$_{12}$-Based Single Molecule Magnet}
\author{E. del Barco$^1$, A. D. Kent$^1$,  N. E. Chakov$^3$, L. N. Zakharov$^2$, A. L. Rheingold$^2$,
D. N. Hendrickson$^2$ and G. Christou$^3$} \affiliation{$^1$Department of Physics, New York University, 4
Washington Place, New York, NY 10003} \affiliation{$^2$Department of Chemistry and Biochemistry, University of
California San Diego, La Jolla, CA 92093-0358} \affiliation{$^3$Department of Chemistry, University of Florida ,
Gainesville, FL 32611-7200}
\date{\today}

\begin{abstract}
A distribution of internal transverse magnetic fields has been observed in single molecule magnet (SMM)
Mn$_{12}$-BrAc in the pure magnetic quantum tunneling (MQT) regime. Magnetic relaxation experiments at 0.4 K are
used to produce a hole in the distribution of transverse fields whose angle and depth depend on the orientation
and amplitude of an applied transverse ``digging field.'' The presence of such transverse magnetic fields can
explain the main features of resonant MQT in this material, including the tunneling rates, the form of the
relaxation and the absence of tunneling selection rules. We propose a model in which the transverse fields
originate from a distribution of tilts of the molecular magnetic easy axes.
\end{abstract}

 \maketitle

Quantum effects in single molecule magnets (SMMs) have been the subject of intense research since their discovery
in the 1990s
\cite{Friedman,Thomas,Hernandez,Sangregorio,Barra,Bokacheva,Hill,delBarco2,Cornia,delBarco3,delBarco4,Hill2}. SMMs
consist of a core of transition metal ions that are strongly exchange coupled with a high spin ferrimagnetic
ground state ($S=4$ to $13$) and uniaxial magnetic anisotropy. The latter leads to a preference for the net spin
to orient parallel or antiparallel to this axis (taken to be the $z$-axis). A number of phenomena have been
clearly observed in these materials, including resonant magnetic quantum tunneling (MQT) between projections of
the spin \cite{Friedman,Thomas,Hernandez}, Berry phase effects \cite{Wernsdorfer} and the crossover between
thermally assisted and pure MQT \cite{Bokacheva}, to name a few. SMMs have also lead experimentalists to develop
more advanced magnetic and spectroscopic techniques to probe the subtleties of MQT. This includes techniques to
examine the distribution of internal axial dipolar or nuclear hyperfine fields \cite{Wernsdorfer2}, molecular
micro-environments \cite{delBarco4} as well as high field single crystal EPR methods \cite{delBarco2,Hill,Hill2}.
The field has also benefited greatly from recently synthesized variations of the original SMM, Mn$_{12}$-acetate,
that are enabling comparative studies of MQT \cite{Christou}. Critical to understanding MQT are methods that
provide access to transverse interactions that break the axial symmetry and thus lead to quantum transitions
between spin-projections.

This can be seen from the form of the effective spin Hamiltonian for SMMs:
\begin{equation}
\label{eq.1}{\cal {H}}=-DS_z^2-BS_z^4-g\mu _B{\bf {H}\cdot {S}}\;,
\end{equation}
The first two terms are the uniaxial magnetic anisotropy of the molecule (positive $D$ and $B$). The third term is
the Zeeman interaction of the spin and the magnetic field. The $2S+1$ allowed projections of the spin, labeled by
$m$, are split by the uniaxial magnetic anisotropy. A longitudinal field, $H_z$, shifts the energy levels favoring
the projections of the magnetization aligned with the field. At well defined longitudinal fields (resonance
fields) the levels $m$ and $m'$ with antiparallel projections on the $z$-axis are nearly degenerate, $H_k\sim
kD/g\mu_B$ ($k=m+m'$). At these resonances MQT is turned on by interactions that break the axial symmetry and mix
the levels $m$ and $m'$, lifting the degeneracy by a small energy, $\Delta_k$, known as the tunnel splitting.
These transverse interactions can be due to transverse magnetic anisotropies (associated with spin-orbit
interactions) and/or a transverse magnetic field, $H_T$.

Mn$_{12}$-BrAc is shorthand for the molecule [Mn$_{12}$O$_{12}$
(O$_{2}$CCH$_{2}$Br)$_{16}$(H$_{2}$O)$_{4}$]-8CH$_{2}$Cl$_{2}$ \cite{An}. The core of the molecule is same as that
of Mn$_{12}$-acetate and the molecule has a $S=10$ ground state. Mn$_{12}$-BrAc molecules have tetragonal
($S_4$-site) symmetry. Hence the lowest order transverse anisotropy term allowed by symmetry is fourth order,
$C(S_+^4+S_-^4)$, and would lead to the tunneling selection rule $m-m'=4i$, with $i$ an integer. Thus pure MQT,
tunneling without thermal activation, is only allowed for resonances $k=4i$. However, in this and all other known
Mn$_{12}$-based SMMs, all tunneling resonances are observed. Further, the tunneling probability increases
monotonically with resonance number $k$. While this means that transverse magnetic fields are present,  up to now
there has been no direct experimental evidence for such fields.

In this Letter, we show clear experimental evidence for a distribution of internal transverse magnetic fields in
Mn$_{12}$BrAc that can explain the MQT phenomena, including the presence of odd-$k$ resonances and the
non-exponential form of the magnetic relaxation. Such transverse fields are likely to be present and important in
other SMMs, including Mn$_{12}$-acetate. However, Mn$_{12}$BrAc is an ideal material to study the effect of
transverse fields because it has small transverse anisotropies, likely because of the nature of the solvent
`micro-environment' around the Mn$_{12}$ core \cite{noteA}.

We have used a high sensitivity micro-Hall effect magnetometer in a low temperature Helium 3 system to measure the
magnetization of a Mn$_{12}$-BrAc single crystal in the pure quantum regime ($T=0.4$ K) where magnetic relaxation
is in the pure quantum tunneling regime \cite{Bokacheva}. Single crystals were removed from the mother liquor just
prior to measurement and immediately placed in grease to minimize solvent loss. Further, x-ray diffraction results
at 100 K on the same crystal used in these experiments show that the crystal is both chemically and
crystallographically the same as that reported previously \cite{An}, but that the solvent (CH$_2$Cl$_2$) content
in the current crystal is double what was previously found. A high field superconducting vector magnet was used to
apply magnetic fields at arbitrary directions with respect to the crystallographic axes of the sample. To study
MQT rates we sweep the applied $z$-axis field at a constant rate $(\nu=dH_z/dt)$ through a resonance $n$ times.
The normalized change in sample magnetization at a resonance, $(M_{before}-M_{after})/(M_{before}-M_{eq})$ (where
$M_{eq}=M_s$), is the MQT probability, $P$ \cite{Wernsdorfer}. For a monodisperse system of molecules this
probability for resonance $k$ is related to the tunnel splitting by the Landau-Zener formula, $P_{LZ}=1-exp\left(
-\frac {\pi\Delta^2}{2\nu_0}\frac n\nu\right)$. Where $\nu_0=g\mu_B(2S-k)$ and $\nu_0\nu$ is the energy sweep
rate. However, for a system with a distribution of tunnel splittings (i.e., due to a distribution of transverse
anisotropy parameters or transverse fields) each molecule in the initial state $m=10$ prior to crossing a
resonance has a different tunneling probability. Therefore, it is possible to study different parts of the
distribution of tunnel splittings in the crystal by appropriate preparation of the initial magnetization state.
\begin{figure}
\begin{center}\includegraphics[width=7cm]{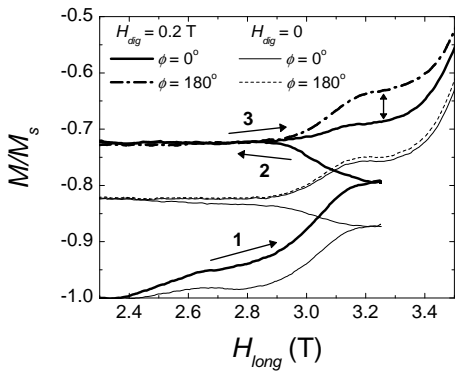}
\vspace{-3 mm} \caption{Hole digging method. Magnetization plotted versus applied longitudinal field. Steps 1 and
2: a transverse digging field, $H_{dig}=$ 0 (thin lines) $H_{dig}=$ 0.2 T (thick lines), is applied along
$\phi_{dig}=0$ during the crossing of resonance $k=6$ back and forth. Step 3: resonance $k=6$ is crossed after the
hole digging process in the presence of a transverse field, $H_T=0.2$ T applied along $\phi=0$ and $180^o$ for
both cases.}\vspace{-9 mm}
\end{center}
\end{figure}

We have developed a hole digging method that allows us to produce a hole in the distribution of transverse fields.
The dependence of the tunnel splitting of a molecule at a resonance $k$ on the transverse field can be
approximated by a power law, $\Delta_k=a_kH_T^{b_k}$, where $a_k$ and $b_k$ are constants that depend on
parameters in the spin Hamiltonian \cite{Garanin}. Consequently, a distribution of transverse fields generates a
distribution of tunnel splittings. In the presence of an external transverse field molecules sensing a given
magnitude and orientation of the internal transverse field will behave in a manner that depends on the magnitude
of the vector sum of these fields. In our method we prepare the initial magnetization state of the system
$M_{ini}=-M_s$. Then we sweep the longitudinal field from 0 T to 3.25 T at $\nu=0.4$ T/min and go back to 0 T
crossing resonance $k=6$ twice in the presence of a digging transverse field, $H_{dig}$, applied at a digging
angle, $\phi_{dig}$ in the $x-y$ plane (steps 1 and 2 in fig. 1). The molecules with highest splitting values have
the greatest probability to relax by MQT. In the next step (step 3 in fig. 1) we sweep the longitudinal field
across the same resonance in the presence of a transverse field, $H_T$, applied in a direction, $\phi$, with
respect to one of the faces of the crystal. In this crossing only those molecules that remained in the metastable
well during the digging process can tunnel across the barrier and contribute to the MQT probability. We repeat
this procedure for different values of $\phi$ from $\phi=\phi_{dig}-180^o$ to $\phi=\phi_{dig}+180^o$. In fig. 1
we show results for two different digging fields, $H_{dig}=$ 0 (dashed line) and 0.2 T (solid line), applied along
$\phi_{dig}=0$ during steps 1 and 2. The measurement in step 3 was done with a transverse field $H_T=0.2$ T
applied along $\phi=0$ and $180^o$ in both cases.

We have carried out hole digging experiments in two different forms. (a) In the first case the values of the
transverse digging field and the transverse field used in the measurement of MQT probability are equal,
$H_{dig}=H_T$. In fig. 2A we show the results obtained with the digging transverse field applied along
$\phi_{dig}=0$ for different values of $H_{dig}=H_T=$ 0.2, 0.25, 0.35 and 0.37 T. (b) In the second case we used a
constant value of the transverse field, $H_T=0.2$ T, in the experiment while we produce the hole with different
transverse digging fields, $H_{dig}=$ 0, 0.025, 0.1, 0.2 and 0.4 T, applied along $\phi_{dig}=0$. The results are
shown in fig. 3A. The MQT probability clearly shows a hole at the same angle, $\phi=0$, for which the digging
transverse field was applied \cite{note1}. This observation unambiguously establishes the existence of a
distribution of transverse fields.

These figures show that: (1) the hole width and depth increase with the magnitude of the digging field; (2) the
probability far from the hole (i.e. $\pm 180^o$) in fig. 3A first increases with the digging field (from
$H_{dig}=0$ to 0.2 T) and then decreases; and (3) the flat response (within the noise) of the curve of $H_{dig}=0$
(fig. 3A) is indicative of the absence of second and fourth order anisotropy terms in the Hamiltonian, which would
lead to 2-fold and 4-fold patterns of maxima, respectively \cite{note2}. This can also be seen in the pink data of
fig. 2A which is insensitive to the orientation of the transverse magnetic field \cite{note3}. This data has been
obtained without using the hole digging process, so in this case all the molecules contribute to the MQT
probability at resonance $k=6$. Observation (2), the fact that P$_{k=6}$($\pm 180^o$) first increases then
decreases with digging field, implies that there are significant transverse fields present in the sample ($\sim$
0.2 T), as will be discussed below.
\begin{figure} \vspace{-2 mm}
\begin{center}\includegraphics[width=7cm]{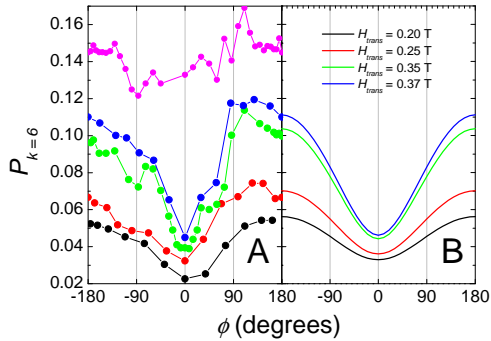}
\vspace{-2 mm}\caption{Hole digging measurements. (A) Measured MQT probability for $k=6$ as a function of the
angle and magnitude of the transverse field and (B) calculation of MQT probability of resonance $k=6$ for
different transverse fields $H_T=H_{dig}$ at $\nu=$ 0.4 T/min., as described in the text.} \vspace{-10 mm}
\end{center}
\end{figure}

To study the behavior of different parts of the distribution of quantum splittings as a function of an external
transverse field we have measured the MQT probability in Landau-Zener multi-crossing measurements of resonance
$k=$ 6 at a sweep rate, $\nu=0.6$ T/min, in the presence of different transverse fields, $H_T=$ 0, 0.1, 0.2, 0.3
and 0.5 T (see ref. \cite{delBarco3} for details on the procedure). The results are shown in fig. 4 starting from
a saturated sample. The fact that relaxation is broad on a logarithm scale confirms the presence of a distribution
of tunnel splittings in the crystal. An interesting result is that the relaxation curves are rather
insensitive to small values of the applied transverse field ($H_T < 0.2$ T).

We can model this data assuming that the origin of the tunnel splitting distribution is a distribution of tilts of
the easy axes of the molecules of a Gaussian form, $f(\alpha,\beta)=exp(-2\alpha^2/\sigma^2)$, where $\alpha$ is
the angle between the easy axis of a molecule and the $c$-axis of the crystal and $\beta$ is the angle between the
$x-y$ plane projection of the easy axis and one of the faces of the crystal (we assume the distribution is
isotropic in the $x-y$ plane). Due to the presence of a longitudinal field, $H_L$, applied along the $c$-axis of
the crystal (i.e. $H_L\sim 3$ T for resonance $k=6$) molecules with different tilts experience different
transverse magnetic fields. When an external transverse field, $H_T$, is applied at an angle $\phi$ with respect
to one of the faces of the crystal the total transverse field felt by a molecule with a tilt $\{\alpha,\beta\}$ is
$h_{T}(\alpha,\beta)\simeq((H_Tcos(\phi-\beta)-H_Lsin\alpha)^2+(H_Tsin(\phi-\beta))^2)^{1/2}$. As mentioned above
the dependence of the tunnel splitting on the transverse field can be approximated by $\Delta_k=a_k h_T^{b_k}$, so
it can be written in terms of $\alpha$ and $\beta$. This approximation is valid for transverse fields below 1 T.
The solution of the Hamiltonian of eq. (1) using $D=$ 0.63 K, $B=$ 1.0 mK and $C=$ 0.023 mK (these parameters have
been extracted from EPR measurements \cite{Hill3}) gives $a_6=1.55\times10^{-5}$ K/T and $b_6=2.06$. To calculate
the MQT probability we have integrated the Landau-Zener probability over the distribution of tilts, where $a_k$,
$b_k$ and $\sigma$ have been used as fit parameters. The results of this calculation are shown in fig. 4
(continuous lines). The fit has been obtained with $a_k=1.5\times10^{-6}$ K, $b_k=3.3$ and $\sigma=14.6^o$. The
values of $a_k$ and $b_k$ are similar to those obtained by diagonalization of the Hamiltonian. Note that the
distribution of transverse fields must be broad to explain the slow increase in the relaxation as a function of
external transverse field, $H_T$, observed in fig. 4. A small change in the width of the tilts distribution (i.e.
$\pm$ 0.2 degrees) leads to a lower quality fit to the data. The distribution of tilts given by
$\sigma=14.6^o$ has a half width at half height (HWHH) of $\alpha_{HWHH}\sim 9$ degrees, which corresponds to a
transverse field of $\sim$0.45 T. We want to point out that our model does not take into account another sources
of disorder (i.e. $D-$ and/or $g-$strain \cite{Park}) that would likely decrease the width of the tilts
distribution necessary to explain the data. On the other hand, the distribution of tilts extracted from our
analysis is sufficient to explain the width of the MQT resonances in the magnetic hysteresis curves,
$H_{L_{HWHH}}\sim 0.1$T.
\begin{figure}
\begin{center}\includegraphics[width=7cm]{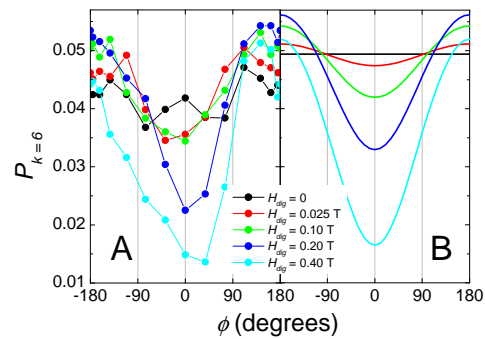}
\vspace{-2 mm}\caption{Hole digging measurements. (A) Measured MQT probability for $k=6$ as a function of the
angle of the transverse field (H$_T=0.2$ T) and magnitude of the digging field $H_{dig}$ (B) Calculation of the
probability as described in the text.}\vspace{-10 mm}
\end{center}
\end{figure}

On the right hand margin of fig. 4 we show 3-d plots that illustrate the portion of molecules within the tilts
distribution that relaxes in the first crossing of resonance $k=6$ in the presence of several external transverse
fields. In figs. 2B and 3B we show the model calculations corresponding to the situation in the experiments in
figs. 2A and 3A respectively, using the parameters obtained from fitting of the multi-crossing relaxation curves.
The agreement with the experiment is excellent. The calculations using the model of a distribution of tilts
reproduce the behavior of the MQT probability close and far from the hole in the distribution. As in the
experiment, one can see in fig. 3B how the probability far from the hole (i.e., $\pm 180^o$) first increases with
the digging field (from $H_{dig}=$0 to 0.2 T) and then decreases. This can be understood from the illustrations in
fig. 4. For low fields the portion of molecules relaxing that are `tilted' antiparallel to the digging field is
significant and consequently the MQT probability far from the hole becomes smaller because some molecules with
this tilt direction relax during the digging procedure. As the digging field is increased, the portion of such
molecules decreases and therefore the MQT probability for $180^o$ increases. The field at which the probability,
P$_{k=6}$($\pm 180^o$), begins to decrease with increasing digging field is the characteristic field scale of
internal transverse field distribution.
\begin{figure}
\begin{center}\includegraphics[width=7cm]{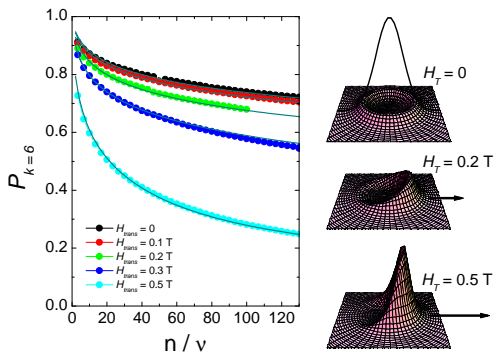}
\vspace{-3 mm} \caption{Multi-crossing relaxation curves of resonance $k=6$ for different applied transverse
fields. The illustrations represent the portion of the tilts distribution that relaxes at $n=1$.} \vspace{-8 mm}
\end{center}
\end{figure}

To summarize, tilts of the easy axis of the molecules may explain the observed broad distribution of transverse
magnetic fields. In this model, small tilts of the easy axis generate a large transverse field due to the high
value of the longitudinal field used in the experiments. Our experimental results also constrain other models of the origin of
the distribution of internal transverse fields. One important thing to point out is that while there are large
transverse fields, there is not a broad distribution of longitudinal fields in the sample (the latter would
completely smooth out the resonances in the magnetization hysteresis loops). This excludes any source of magnetic
fields that are randomly oriented (i.e. dipolar or hyperfine fields). Also, the magnitude of the observed
transverse fields precludes dipolar and hyperfine fields  ($\leq 0.05$ T). Tilts of the magnetic easy axis
were first shown to occur due to crystal dislocations \cite{Chudnovsky}. We also note that large tilts
have been observed in a low anisotropy barrier form of Mn$_{12}$ and associated with an unusual distortion
around a Mn(III) site on the outer part of the molecule \cite{Takeda}. Exchange interaction are another possible
origin of internal fields, but at present, there is no evidence for such interactions in Mn$_{12}$-BrAc. Tilts and
distributions of internal transverse fields may be important in explaining MQT relaxation at odd resonances
observed in other studied SMMs (i.e., Mn$_{12}$-acetate, Fe$_8$,...). The implementation of this method to probe
transverse fields in other SMMs and ensembles of nanomagnets will further our understanding of MQT.

The authors acknowledge useful discussions with S. Hill and N. Dalal. This research was supported by NSF.


\begin{references}
\bibitem{noteA} In Mn$_{12}$-acetate there are on average 2 acetic acid (CH$_3$COOH) solvents per molecule and
4 possible solvent positions around the molecule. Disorder in the position of these solvents lowers the molecule
symmetry and leads to a strong second order transverse anisotropy \cite{Hill2,delBarco4}. In Mn$_{12}$-BrAc there
are 8CH$_{2}$Cl$_{2}$ solvents/mol. and weaker solvent/molecule hydrogen bonding interactions and
no apparent second order transverse anisotropy \cite{Hill3}.
\bibitem{note1} We have done hole digging experiments by applying the digging transverse field at
an angle, $\phi_{dig}=200^o$. The result showed a hole equivalent to the measurements shown in this Letter but
centered at $\phi=200^o$. This constitutes evidence that the hole orientation depends on the orientation of the
digging field and supports our assumption of an isotropic distribution of tilts in the $x-y$ plane.
\bibitem{note2} For the amplitudes of the transverse magnetic fields applied in these experiments
($\leq 0.4$ T) the expected change in tunneling probability with angle $\phi$ is negligible on the scale of the
plots in Fig. 2 and 3 \cite{delBarco3}. However, the 4-fold rotation symmetry in the $x-y$ plane has been seen in
recent EPR experiments in which larger transverse magnetic fields ($\sim 4$ T) are employed \cite{Hill3}.
\bibitem{note3} The minimum and the maximum observed at $\phi=\pm90^o$ in the pink curve of fig. 2A
are due to a small non-linearities in the Hall magnetometer when there is a ($\sim 0.3$ T) magnetic field applied
perpendicular to the plane of the sensor. This also leads to the fluctuations seen in the other curves in fig 3A.
\end{references}
\end{document}